\documentstyle[twocolumn,aps,epsfig]{revtex} 

\begin{document}
\twocolumn[\hsize\textwidth\columnwidth\hsize\csname @twocolumnfalse\endcsname

\title{Thermodynamics of the anisotropic Heisenberg chain calculated 
by the density matrix renormalization group method}
\author{Naokazu Shibata}
\address{Institute for Solid State Physics, University of Tokyo,  
7-22-1 Roppongi, Minato-ku, Tokyo 106, Japan}
%\recdate{\today}

\maketitle

\begin{abstract}
The density matrix renormalization group (DMRG) method is applied 
to the anisotropic Heisenberg chain at finite temperatures. 
The free energy of the system is obtained using the quantum transfer 
matrix which is iteratively enlarged in the 
imaginary time direction. The magnetic susceptibility and the specific 
heat are calculated down to $T=0.01J$ and compared with the 
Bethe ansatz results. The agreement including the 
logarithmic correction in the magnetic susceptibility at the isotropic 
point is fairly good.
\end{abstract}

\vskip2pc]

%\kword{density matrix, renormalization group, Heisenberg model,
%anisotropy, magnetic susceptibility, specific heat}
%\def\runtitle{Thermodynamics of the anisotropic Heisenberg chain}
%\def\runauthor{N. Shibata}

\narrowtext

%\sloppy
The density matrix renormalization group (DMRG) method has 
brought significant progress to the field of numerical calculations. 
This method provides a way to study 
long chains by iteratively enlarging the 
system size and to obtain the ground state wave function 
with only small systematic errors~\cite{DMRG}. 
Although the DMRG method was initially designed for 
one-dimensional quantum systems it has also been applied 
to two-dimensional classical systems~\cite{Nishino},
and now even to two-dimensional quantum systems~\cite{SW}.
A recent advance in the use of the DMRG method is its 
application to thermodynamics~\cite{Trans,FTRG}. 
In the present paper, we apply the DMRG method to the 
quantum transfer matrix and obtain accurate results which 
allow us to study low temperature behaviors of 
thermodynamic quantities such as the logarithmic correction
in the magnetic susceptibility of the Heisenberg chain.

Initially, the DMRG procedure at finite temperature 
was designed to be a similar algorithm to the zero 
temperature DMRG except that in the former we take several relevant 
low-energy states as target states to obtain a representation
of the Hamiltonian most efficiently at a given temperature~\cite{DMRG,FTRG}. 
By adding the density matrices, calculated from the 
low-energy states using a normalized Boltzmann weight,
one can construct an effective density matrix and 
find important bases by diagonalizing it.
While this method was successfully applied to finite systems,
its application to the infinite system 
is difficult because, in general, we need to increase the size 
of the system while decreasing the temperature, 
which introduces an enormous truncation error to the calculation.

To deal with the infinite system in a more systematic way,
one can use the transfer matrix method.
In this method the free energy of the infinite system is 
directly obtained from the maximum eigenvalue of the 
transfer matrix~\cite{Bets,Koma}.
The partition function of the system whose Hamiltonian $H$ is 
decomposed into two parts $H_{\mbox{\scriptsize odd}}=
\sum_{n=1}^{L/2}h_{2n-1,2n}$ and 
$H_{\mbox{\scriptsize even}}=\sum_{n=1}^{L/2}h_{2n,2n+1}$ such as 
$[h_{2n-1,2n},h_{2n'-1,2n'}]=[h_{2n,2n+1},h_{2n',2n'+1}]=0$
is represented by the transfer matrix ${\cal T}$ as
\begin{eqnarray}
Z & = & \mbox{\boldmath Tr}\ \mbox{\boldmath e}^{-\beta H} \\ 
  & = & \lim_{M \rightarrow \infty}  \mbox{Tr}\  
        [ ( \mbox{\boldmath e}^{-\beta H_{\mbox{\tiny odd}}/M} 
              \mbox{\boldmath e}^{-\beta H_{\mbox{\tiny even}}/M} ) ]^M \\
  & = & \mbox{\boldmath Tr}\ \prod_{n=1}^{L/2}{\cal T}_n 
\end{eqnarray}
where  ${\cal T}_n$ is defined by
\begin{eqnarray}
{\cal T}_n & = & \lim_{M \rightarrow \infty} 
[\mbox{\boldmath e}^{-\beta h_{2n-1,2n}/M} 
\mbox{\boldmath e}^{-\beta h_{2n,2n+1}/M} ]^M .
\end{eqnarray}
Thus, once we know the maximum eigenvalue 
$\lambda_{\mbox{\scriptsize max}}$ of ${\cal T}_n$,
we can get the partition function 
\begin{eqnarray}
Z & = & \lambda_{\mbox{\scriptsize max}}^{L/2}
\end{eqnarray}
when ${\cal T}_n$ is translationally invariant and 
the length of the system $L$ is large enough.

The problem now, is how to calculate
$\lambda_{\mbox{\scriptsize max}}$ 
of ${\cal T}$, the matrix size of which diverges with 
increasing $M$. To manage this difficulty we use the DMRG 
method~\cite{Trans}:
We iteratively increase $M$ in ${\cal T}$ by
restricting the number of bases used in 
${\cal T}$ and keeping $\lambda_{\mbox{\scriptsize max}}$ 
optimal.
For this purpose we define $ {\cal T}^A(M=2) =
\mbox{\boldmath e}^{-\beta_0 h_{\mbox{\tiny odd}}} $ and $ {\cal T}^B(M=2)
= \mbox{\boldmath e}^{-\beta_0 h_{\mbox{\tiny even}}} $ with 
$h_{\mbox{\scriptsize odd}}=h_{2n-1,2n}$
and $h_{\mbox{\scriptsize even}}=h_{2n,2n+1}$,
and increase $M$ of 
${\cal T}^{A (B)}(M)$ as
\begin{eqnarray}
 {\cal T}^A(M) \mbox{\boldmath e}^{-\beta_0 h_{\mbox{\tiny even}}} 
& \rightarrow & {\cal T}^{A'}(M+1) \\
 \mbox{\boldmath e}^{-\beta_0 h_{\mbox{\tiny odd}}} {\cal T}^B(M) 
& \rightarrow & {\cal T}^{B'}(M+1) ,\\
{\cal T}^{A'}(M) \mbox{\boldmath e}^{-\beta_0 h_{\mbox{\tiny odd}}} 
& \rightarrow & {\cal T}^{A}(M+1) \\ 
 \mbox{\boldmath e}^{-\beta_0 h_{\mbox{\tiny even}}} {\cal T}^{B'}(M) 
& \rightarrow & {\cal T}^{B}(M+1) .
\end{eqnarray}
The optimal set of bases 
at the transformations in eq.~(6)
will be determined as follows.
We first diagonalize 
${\cal T}={\cal T}^{A} \mbox{\boldmath e}^{-\beta_0 h_{\mbox{\tiny even}}} 
\mbox{\boldmath e}^{-\beta_0 h_{\mbox{\tiny odd}}} {\cal T}^{B}$ 
with the periodic boundary conditions in the $\beta$ direction,
and obtain left and right eigenvectors for ${\cal T}$, 
$V^L$ and $V^R$, which correspond to the maximum eigenvalue of ${\cal T}$.
We then compose the density matrix $\rho_{i i'}^A$
for the bases of ${\cal T}^{A} 
\mbox{\boldmath e}^{-\beta_0 h_{\mbox{\tiny even}}}$ as
\begin{eqnarray}
\rho_{i i'}^A = \sum_j V^L_{i j} V^R_{i' j} 
\end{eqnarray}
where $i$ and $j$ are indices of the bases for 
${\cal T}^{A} \mbox{\boldmath e}^{-\beta_0 h_{\mbox{\tiny even}}}$ and 
$\mbox{\boldmath e}^{-\beta_0 h_{\mbox{\tiny odd}}} {\cal T}^{B}$ 
as shown in Fig.~1. The best way to restrict the number of bases 
is to find a proper transformation to new bases 
$i \rightarrow i_{\mbox{\scriptsize new}}$
which maximize the norm $\sum_{i j} V^L_{i j} V^R_{i j}$,
%which should be unity if number of bases is complete,
with a fixed number $m$ of $i_{\mbox{\scriptsize new}}$.
To do this we use the left and right eigenvectors
of the $m$ largest eigenvalues
$\gamma^{i_{\mbox{\scriptsize new}}}$ of the 
$\rho_{i i'}^A$ to be the transformation matrix to
the new bases $i_{\mbox{\scriptsize new}}$.
The error accompanied by this truncation 
is defined by
$1-\sum_{i_{\mbox{\scriptsize new}}=1}^m 
\gamma^{i_{\mbox{\scriptsize new}}}$.
\begin{figure}
%\figureheight{8cm}
\epsfxsize=80mm \epsffile{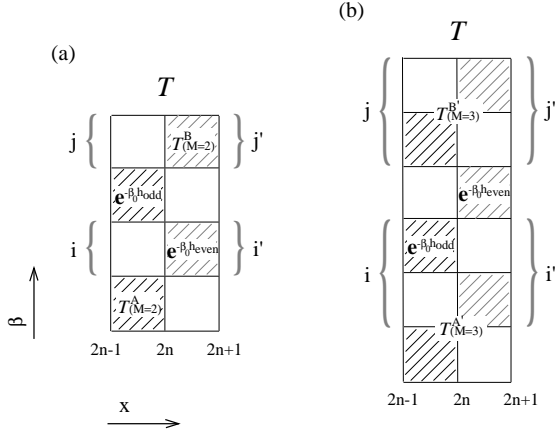}
%\epsfile{file=trans.ps,height=8cm}
\caption{ 
%Fig. 1 \ \
(a) Transfer matrix used at the renormalization in eq.~6. ($M=2$)\ \ 
%\hspace*{1.6cm} 
(b) Transfer matrix used at the renormalization in eq.~8. ($M=3$)
}
%\label{Fig1}
\end{figure}

The most significant difference of the DMRG procedure for the 
transfer matrix from the zero-temperature
DMRG algorithm is that, in the former, we have to use the 
non-symmetric density matrix.
This leads to the left and right eigenvectors of the 
density matrix, $u^{i_{\mbox{\scriptsize new}}}_i$ and 
$v^{i_{\mbox{\scriptsize new}}}_i$, being different. 
For this reason we have to transform the transfer matrix 
using these vectors as
%$v^{i_{\mbox{\scriptsize new}}}_i$ and 
%$u^{i_{\mbox{\scriptsize new}}}_i$ as
\begin{equation}
{\cal T}^{A}_{i_{\mbox{\scriptsize new}} i_{\mbox{\scriptsize new}}'}
 = \sum_{i i'} v^{i_{\mbox{\scriptsize new}}}_i {\cal T}^{A}_{i i'}
u^{i_{\mbox{\scriptsize new}}'}_{i'}
\end{equation}
where $v^{i_{\mbox{\scriptsize new}}}_i$ and 
$u^{i_{\mbox{\scriptsize new}}}_i$ satisfy the orthogonality
\begin{equation}
 \sum_i v^{i_{\mbox{\scriptsize new}}}_i 
 u^{i_{\mbox{\scriptsize new}}'}_i = 
\delta_{i_{\mbox{\scriptsize new}} i_{\mbox{\scriptsize new}}'} .
\end{equation}
Technically, the non-symmetric nature of the density matrix 
is a difficult point of the numerical calculation. 
In fact, if we do not pay attention to the 
number of states kept, the eigenvalue of the density 
matrix becomes complex. 
In order to avoid this we fix the number states
in the density matrix throughout the calculation. 

In the following 
we apply the above DMRG method to the 
S=1/2 anisotropic Heisenberg chain to check the reliability 
of the method. The Hamiltonian is given by
\begin{equation}
H= J \sum_{i} \{ S_i^xS_{i+1}^x+S_i^yS_{i+1}^y+\Delta S_i^zS_{i+1}^z\} ,
\end{equation}
where $\Delta$ is the anisotropy parameter which changes the 
interaction from XY type to Ising type with increasing
$\Delta$: $\Delta=0$ corresponds to the
XY limit, and $\Delta=1$ the isotropic point. 
For this model it is well known that there are three different regimes. 
When $\Delta>1$ a finite gap exists in the excitation spectrum and 
both the magnetic susceptibility and the specific heat show exponential 
decay at low temperatures. At $\Delta=1$, the excitation gap closes, 
and the system shows a critical behavior 
with a logarithmic correction in $\chi$ at low temperatures~\cite{BA}. 
In the region $\Delta<1$, the logarithmic correction in $\chi$ 
disappears and $\chi$ approaches a finite value 
in accordance with a power law. 
Thus this model provides a rich variety of structures in the 
thermodynamic 
quantities and is an appropriate test case for the present method.

In this calculation we use the simple power method to obtain 
the maximum eigenvalue and corresponding eigenvector of the 
non-symmetric transfer matrix, and 
use the lapack library to diagonalize the non-symmetric density matrix. 
The maximum number of states $m$ in the density matrix 
is 76 and the corresponding truncation error is $5\times 10^{-5}$ 
at the lowest temperature $T=0.01J$. In
order to carry out the extrapolation for the Trotter number $M$ we 
start the DMRG procedure from several initial temperatures 
$\beta_0$ and iteratively enlarge the transfer matrix in the 
imaginary time direction using the infinite system algorithm with 
decreasing temperature.

We first show magnetic susceptibility $\chi^z$ obtained by 
the present method. $\chi^z$ is determined from the change in 
the free energy $\delta F=\chi^z h^2/2$ in a small magnetic 
field $h=0.01J$ along the z-axis. The calculated $\chi^z$ in the XY 
limit is shown in Fig.~2(a). The exact result is also plotted in the
figure~\cite{Katsu}.
The agreement with the exact result is excellent even at the 
lowest temperature $T=0.01J$. 
It should be mentioned that such a quadratic behavior in 
$\chi^z$ is difficult to reproduce if we use a simulation 
based on a finite system where 
$\chi^z$ always vanishes exponentially or diverges in the 
low temperature limit.
The result at $\Delta=0.5$ is presented in Fig.~2(b).
Similarly to Fig.~2(a), $\chi^z$ decreases to a finite 
value with a power law with decreasing temperature. 
The obtained power law exponent 
is $2.0$ which is consistent with the result of the 
bosonization~\cite{BA}.
The isotropic case is presented in Fig.~2(c). In this case 
the result of the Bethe ansatz is also plotted for comparison. 
The rather rapid decrease in the low temperature 
region is due to the logarithmic correction. 
The small discrepancy at $T=0.01J$ is considered 
to originate from the finite nature of 
the magnetic field and from truncation errors in the DMRG. 
An Ising like case $(\Delta =2.0)$ is presented in Fig.~2(d). 
In this case $\chi^z$ approaches zero with an exponential form
consistent with the existence of an excitation gap.
The estimated value of the gap is consistent 
with the exact value $0.15J$~\cite{Ising}. 

\begin{figure}
%\figureheight{16cm}
\epsfxsize=80mm \epsffile{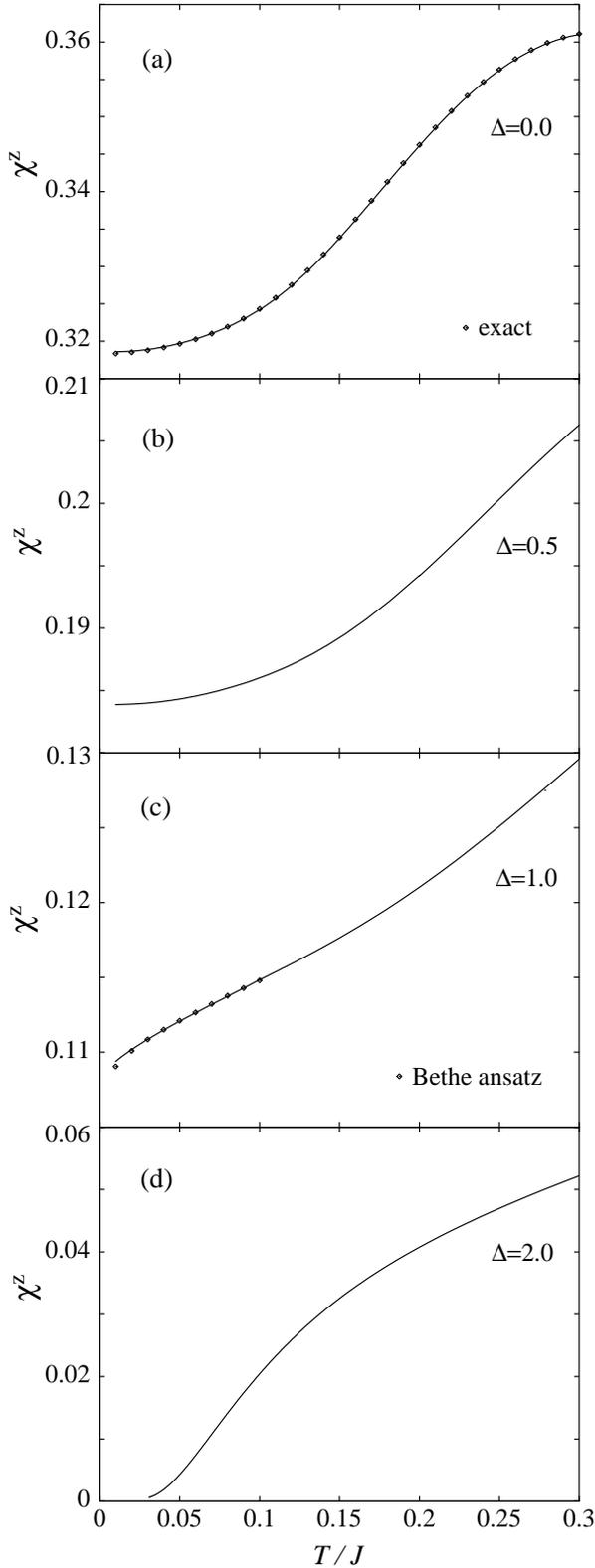}
%\epsfile{file=sus_all2.ps,height=16cm}
\caption{
%Fig. 2 \ \ 
Magnetic susceptibility of the S=1/2 Heisenberg chain
calculated by the DMRG.
}
\label{Fig2}
\end{figure}

We next calculate the specific heat. The specific heat is 
determined by the second 
derivative of the free energy; $C=-T\partial^2 F/\partial T^2$. 
The results for $\Delta=0.0,0.5,1.0$, and $2.0$ are shown in Fig.~3. 
The exact results in the XY limit $(\Delta=0)$ are also plotted. 
In the XY-limit the DMRG results reproduce the exact values well. 
For $\Delta \le 1.0$ the specific heat vanishes linearly as $T$ 
approaches zero.
For $\Delta=2.0$, on the other hand, it seems to be almost zero
at finite $T$ consistent with the existence of the spin gap.
Thus all the results obtained by the 
present DMRG calculation are consistent with the expected behaviors.
\begin{figure}
%\figureheight{6cm}
\epsfxsize=80mm \epsffile{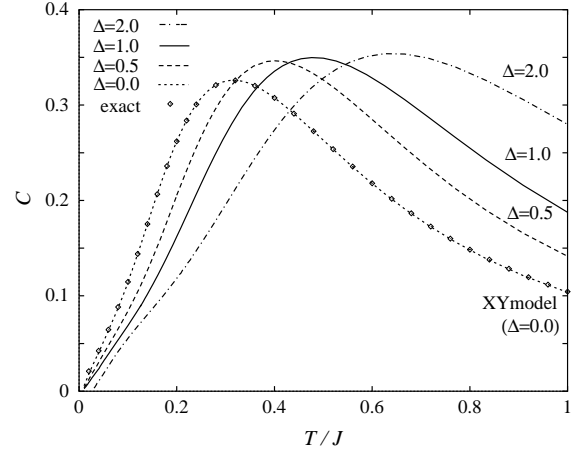}
%\epsfile{file=c_all_ex.eps,height=4.5cm}
\caption{
%Fig. 3 \ \ 
Specific heat of the S=1/2 Heisenberg chain 
calculated by the DMRG
}
\label{Fig3}
\end{figure}

To summarize, we have shown that 
the DMRG method at finite temperature 
is formulated using the transfer matrix 
and that this method actually provides reliable 
results even at low temperatures.
The magnetic susceptibility and the specific heat of the 
Heisenberg chain are accurately calculated
for both cases in which the excitation spectrum is gapless and gapfull.
The logarithmic correction in the magnetic susceptibility is 
correctly reproduced at the isotropic point 
where the system shows the highly critical behavior.
Since this method can be applied to both 
fermion systems and frustrated spin systems,
the DMRG method at finite temperatures
will be a promising numerical method with which to study the low 
temperature properties of one-dimensional quantum systems. 

During the present calculation the author was informed that
X. Wang and T. Xiang have done similar work~\cite{Wang}.
The author would like to thank 
Beat Ammon, Manfred Sigrist, Matthias Troyer,
and Kazuo Ueda for valuable discussions.
Numerical calculations were performed on VPP500
at the ISSP, University of Tokyo.
The author is supported by the Japan Society for the 
Promotion of Science.

\end{document}